# Anomalous Hall effect at the spontaneously electron-doped polar surface of PdCoO$_2$ ultrathin films

T. Harada[1]*, K. Sugawara[2,3,4], K. Fujiwara[1], S. Ito[1], T. Nojima[1], T. Takahashi[2,3,4], T. Sato[2,3,4], and A. Tsukazaki[1,4]

[1]*Institute for Materials Research, Tohoku University, Sendai, Japan*
[2]*Department of Physics, Tohoku University, Sendai, Japan*
[3]*WPI Advanced Institute for Materials Research, Tohoku University, Sendai, Japan*
[4]*Center for Spintronics Research Network, Tohoku University, Sendai, Japan*
*Correspondence to: t.harada@imr.tohoku.ac.jp*

We revealed the electrical transport through surface ferromagnetic states of a nonmagnetic metal PdCoO$_2$. Electronic reconstruction at the Pd-terminated surface of PdCoO$_2$ induces Stoner-like ferromagnetic states, which could lead to spin-related phenomena among the highly conducting electrons in PdCoO$_2$. Fabricating a series of nanometer-thick PdCoO$_2$ thin films, we detected a surface-magnetization-driven anomalous Hall effect via systematic thickness- and termination-dependent measurements. Besides, we discuss that finite magnetic moments in electron doped CoO$_2$ triangular lattices may have given rise to additional unconventional Hall resistance.

Layered transition-metal oxides have given great insight into the role of dimensionality in correlated electron physics [1-4]. Among the many different oxides, the delafossite metal PdCoO$_2$ is a unique system in which monovalent Pd$^+$ ions are stabilized in the form of a two-dimensional (2D) Pd sheet that is sandwiched by [CoO$_2$]$^-$ layers [5-7] (Fig. 1 (a)). This 2D Pd$^+$ sheet (Fig. 1 (b)) provides a highly dispersive 4$d$-dominant conduction band [8-11], while [CoO$_2$]$^-$ forms an insulating block layer (Fig. 1 (c)). PdCoO$_2$ bulk single crystals exhibit highly anisotropic electrical conduction, with the electrical conductivity exceeding that of elemental Pd [6]. More surprisingly, the long mean free path of the interacting dense electrons (~ 20 μm) induces hydrodynamic collective electron motion, as observed in high-mobility semiconductor heterostructures, graphene, and tungsten phosphide [12-16].

Because of the ionicity of the Pd$^+$ and [CoO$_2$]$^-$ layers (Fig. 1(a)), spontaneous charge compensation is induced at the polar surface. The electronic structure of the surface is expected to be significantly different from that of the bulk, and to show a strong termination-dependence. In particular, surface ferromagnetic state emerges at the Pd-terminated surfaces [17]. The surface charge compensation introduces extra electrons into the Pd surface layer, and this increase in electron density shifts the surface-Pd energy band to a higher binding energy relative to the original bulk energy band. This results in a flat branch of the surface band occurring near the Fermi level ($E_F$). The high density of correlated electrons in the flat band increases the Stoner instability, resulting in the formation of spin-split surface Pd bands [17], as shown schematically in Fig. 1(d). This causes the Pd-terminated surface to become ferromagnetic, even though bulk PdCoO$_2$ is nonmagnetic [17].

Indeed, Stoner-like splitting has been detected in the Pd-terminated surface region of PdCoO$_2$ bulk single crystals using angle-resolved photoemission spectroscopy (ARPES) [18]. This observation suggested that a polar surface that contained highly conducting and spin-polarized electrons existed, which may lead to exotic spin-related transport phenomena in nonmagnetic PdCoO$_2$. However, it is difficult to probe this type of surface transport in bulk samples because current shunting by the relatively vast bulk volume obscures the surface conduction. To overcome this, here we examine ultrathin PdCoO$_2$ films. By altering film thickness [19], we have succeeded in detecting the signal from spin-related transport in these ultrathin films, demonstrating that the surface ferromagnetism occurred within a limited thickness $d_s$ near the surface (Fig. 1(e) and Fig. S1[20]).

The preparation of Pd-terminated surfaces is essential to induce surface ferromagnetism via Stoner splitting [17]. We adopted a thin-film approach in this study, as opposed to the bulk cleavage method. We fabricated nanometer-thick PdCoO$_2$ films on Al$_2$O$_3$ (0001) ($c$-Al$_2$O$_3$) substrates using pulsed laser deposition (see Supplemental Material [20] for detailed methods). The $c$-axis oriented growth of PdCoO$_2$ (Fig. 1(a)) was confirmed using X-ray diffraction (XRD) (Fig. S2 [20]), as reported in detail previously [19]. A typical, high-angle annular dark-field scanning transmission electron microscope (HAADF-STEM) image around the film-substrate interface is shown in Fig. 1(f). The bright and dark layers clearly alternated along the normal to the film plane, which correspond to the Pd$^+$ and [CoO$_2$]$^-$ layers, respectively. Because the initial growth layer was [CoO$_2$]$^-$, and to ensure charge neutrality of the whole thin film, the final layer (top surface) was likely to be Pd$^+$, although this could not be resolved in HAADF-STEM image. The electronic band dispersion at the film surface was characterized directly using ARPES as a surface-sensitive probe [18,21,22]. A pair of highly dispersive bands (denoted as $\alpha$ and $\beta$) were observed along the Γ-K cut (Fig. 1(g)), which was consistent with Pd 4$d$-derived states [18]. The fact that there were no traces of Co-derived hole-like states (region within the dotted purple square in Fig. 1(g)), combined with the Fermi surface map (Fig. 1(h)), was consistent with the Pd-terminated surface that was observed for the bulk single crystal [18]. As such, the ARPES spectra indicated that the final layer in the PdCoO$_2$ films that were fabricated in this study was predominately Pd. Furthermore, the non-degenerated $\alpha$ and $\beta$ bands and the flat branch of the $\alpha$ band located just below $E_F$ (Fig. 1(g)) are all hallmarks of Stoner-like splitting [18].

The Hall resistance ($R_H$) versus magnetic field ($\mu_0 H$) curve for a 13-nm-thick PdCoO$_2$ film at 2 K is shown in Fig. 2(a). The $R_H$ curves were obtained by subtracting the $\mu_0 H$-linear ordinary term, determined by the high-field data ($\mu_0 H$ = 5–7 T), from the measured transverse resistance ($R_{yx} = V_{yx} / I_x$) (Fig. S3 [20]). While only a small nonlinear signal was observed from the 13-nm-thick sample (Fig. 2(a)), the signal became more pronounced as the thickness ($d$) of the $PdCoO_2$ film was decreased, as shown in Figs. 2(b) and 2(c). The comparable ordinary Hall coefficients from the three samples (the linear gradient at $\mu_0 H$ = 5 – 7 T in Figs. S3(a)-(c) [20]) indicated that the large $R_H$ was induced by a reduction in $d$, rather than change in the thin-film quality.

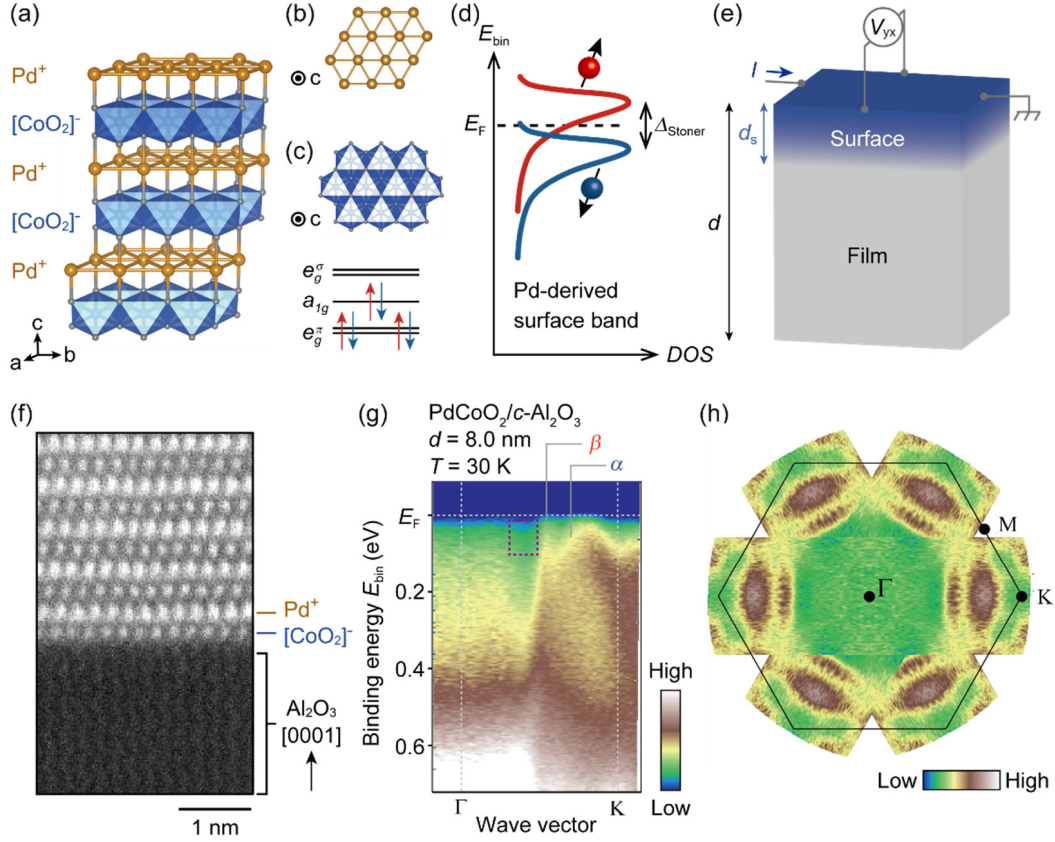

**FIG. 1.** Crystal structure and electronic structure of $PdCoO_2$. (a) Crystal structure of $PdCoO_2$. (b) Two-dimensional Pd layer. (c) $CoO_2$ sublattice (top) and $Co^{3+}$ spin arrangement (bottom). (d) The density of states (DOS) of surface Pd bands split by $\Delta_{Stoner}$ by Stoner instability [17,18]. (e) Schematic illustration of a $PdCoO_2$ thin film with a surface ferromagnetic state extending over $d_s$. Hall effect measurement probes the conduction throughout the entire film. (f) HAADF-STEM image of a $PdCoO_2$ thin film on $Al_2O_3$ (0001), projected along the $Al_2O_3$ $[\bar{1}100]$ direction. Pd and $CoO_2$ sublattices can be seen on the $Al_2O_3$ substrate. (g) The electronic band dispersion along the Γ-K cut of a $PdCoO_2$ film with $d$ = 8.0 nm. The spin-split surface Pd bands are shown as $\alpha$ and $\beta$. The purple dotted square indicates the region where the Co-derived surface band is expected to appear if it existed [18,22]. (h) Fermi surface of $PdCoO_2$ obtained by symmetrizing the ARPES data in (g).

A log-log plot of the sheet resistance under zero field ($R_{sheet}$) as a function of $d$ is shown in Fig. 2(d). Fitting this plot yielded $d\log(R_{sheet}) / d\log(d)$ of $-1 \pm 0.2$ (blue line in Fig. 2(d)), and so $R_{sheet}$ was inversely proportional to $d$. Therefore, the three-dimensional electrical conductivity (S cm$^{-1}$) was constant, which secured that the quality of the films was unchanged down to $d = \sim 3$ nm. In contrast, fitting the plot of $R_H$ at 9 T ($R_H^{9T}$) versus $d$ (Fig. 2(e)) resulted in $d\log(R_{sheet}) / d\log(d)$ of $-2 \pm 0.3$ (red line in Fig. 2(e)). The inverse square dependence of $R_H^{9T}$ on $d$ ($d^{-2}$) can be understood by assuming that only the finite region near the surface was generating the nonlinear Hall resistance (See Supplemental Material for the detailed model [20]). This was consistent with the surface charge compensation model [17].

The magnetism of the $PdCoO_2$ ultrathin films was examined using a superconducting quantum interference device. The saturation magnetization at 7 T ($M_s^{7T}$) was plotted as a function of temperature, as shown in Fig. 2(f). The $M_s^{7T}$ was estimated from the magnetization versus $\mu_0 H$ curves by subtracting the background signal from the $c$-$Al_2O_3$ substrate (Fig. S4(a) [20]). Below approximately 30 K, the $M_s^{7T}$ gradually increased as the temperature was decreased, following the Bloch formula [23] that is typically applied to ferromagnets. In addition, the $M_s^{7T}$ was independent of $d$ (Fig. S5 [20]), and therefore the magnetized volume in all the films was nearly constant. As shown in Fig. 2(f), both $R_H^{9T}$ and $M_s^{7T}$ exhibited similar temperature dependence. These observations indicated that the anomalous Hall effect (AHE) that was observed in the Pd-terminated ferromagnetic surface was responsible for the nonlinear Hall resistance.

According to the Stoner splitting model, surface magnetism should vanish when the surface polarity is cancelled. Taking advantage of the thin-film growth technique, we attempted to modulate the surface polarity by over layer deposition using the (111)-surface of rock-salt-type CoO, a single layer of which mimics the $[CoO_2]^-$ layer in $PdCoO_2$ (Fig. 3(a)). After the growth of $PdCoO_2$, a 5-nm-thick $CoO_x$ layer was deposited. The growth conditions used were identical to those used for $PdCoO_2$ growth to minimize any influence on the $PdCoO_2$ film quality during the $CoO_x$ deposition. Although the exact phase of the thin $CoO_x$ capping layer was not identified from the XRD measurements, we consider the CoO (111) / PdCoO$_2$ (0001) heterointerface to be the ideal model for the CoO$_x$-capped PdCoO$_2$ (Fig. 3(a)). Capping with CoO$_x$ drastically reduced the $R_H$ by nearly 45% of the original value, while maintaining the saturation field (Fig. 3(b)). The remaining finite $R_H$ may have been caused by an electronic reconstruction that was required to satisfy charge neutrality throughout the entire PdCoO$_2$ layer. Although further characterization is needed for quantitative discussions, this result underpins the Stoner origin of the observed AHE.

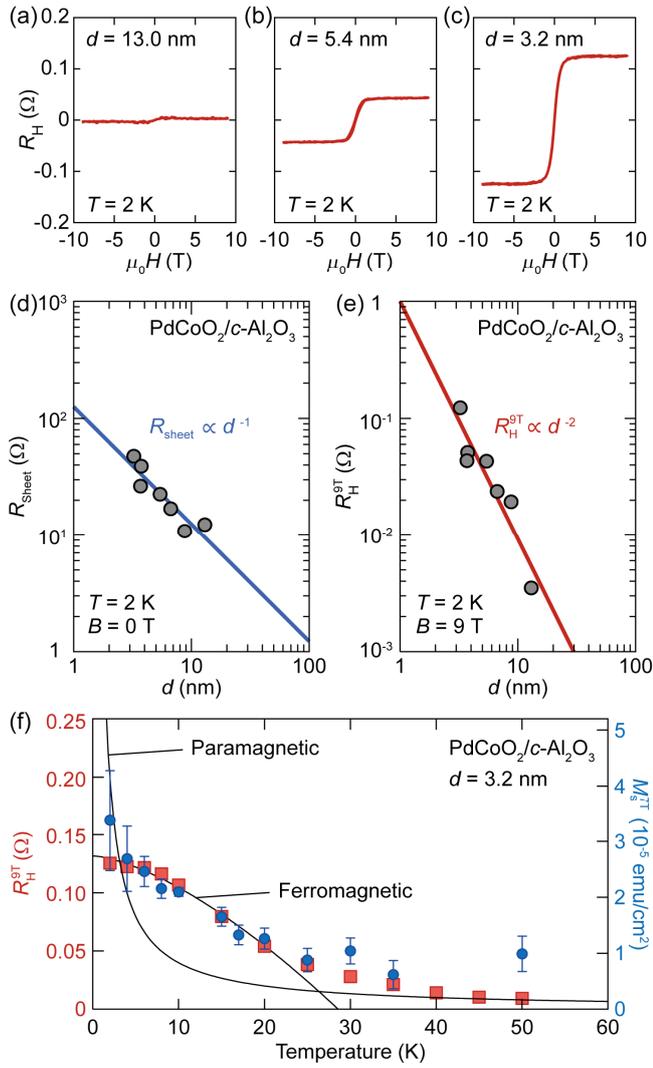

FIG. 2. Anomalous Hall effect of PdCoO$_2$ thin films. (a)–(c) $R_H$-$\mu_0 H$ curves of PdCoO$_2$ films measured at 2 K with different thicknesses ($d$) ($d$ = 13.0 nm, 5.4 nm, and 3.2 nm, respectively). (d) Sheet resistance ($R_{sheet}$) at $B$ = 0 T and (e) $R_H$ at $B$ = 9 T ($R_H^{9T}$) as functions of $d$, plotted as log-log graphs (the red and blue lines are fitting results). (f) Temperature dependence of $R_{AH}^{9T}$ (red squares) and the saturation magnetization measured at $\mu_0 H$ = 7 T ($M_s^{7T}$) (blue circles) for a film with $d$ = 3.2 nm. The black curves correspond to typical ferromagnetic and paramagnetic behaviors.

Comparison of the $R_H$ and magnetization ($M$) data highlighted a discrepancy in their behavior at low fields (Figs. S4(a) and S4(b) [20]). Changes in the measured $R_H$ and the magnetization-driven anomalous Hall (AH) resistance with temperature are shown in Fig. 4(a). The $R_{AH}$-$\mu_0 H$ curve (red) was calculated from the $M$-$\mu_0 H$, by assuming that $R_{AH} \propto M$. As highlighted in blue, their difference (i.e. $R_{UH} = R_H - R_{AH}$) became significant as the temperature was lowered. When calculating $R_{UH}$, we assumed that the $R_H$ at 7 T was dominated by the magnetization-driven $R_{AH}$ because the high-field saturation values of $R_{AH}$ and $M$ as functions of temperature overlapped (Fig. 2(f)). This assumption corresponds to the situation in which the spins are parallel to the magnetic field at 7 T. The $R_{UH}$ (peak value) and $R_{AH}$ each exhibited a distinct temperature dependence, which indicated that they had different origins (Fig. 4(b)). The map of $R_{UH}$ in the temperature-$\mu_0 H$ plane (Fig. 4(c)) revealed a dome-like structure that peaked at low field ($\mu_0 H < 3$ T). As mentioned previously, the $d^{-2}$ dependence of $R_{UH}$ indicated that $R_{UH}$ originated on the surface (Fig. S6 [20]).

The origin of the $R_{UH}$ is discussed by considering the role of the CoO$_2$ layer near the surface. The characteristic peaks that were observed at low field strength (1–2 T) in the $R_{UH}$ plots in Fig. 4(a) were similar to the unconventional anomalous Hall effect reported for systems with non-uniform spin textures [24-28]. Nonlinear Hall resistance occurs in antiferromagnetic PdCrO$_2$, which is a structural analogue of PdCoO$_2$, despite the antiferromagnetic magnetization versus $\mu_0 H$ curve [26]. The spin-frustrated triangular Cr$^{3+}$ lattice is thought to play a role via the non-colinear spin arrangement [26]. In nominal PdCoO$_2$ that contains Co$^{3+}$ ions in the nonmagnetic low-spin state (Fig. 1(c)) [9,29], this mechanism is not expected to occur. However, when extra electrons are partially doped into the [CoO$_2$]$^-$ layer, the resultant finite moments within the Co$^{2+}$ ions (Fig. 4(d)) could locally form a similar non-colinear spin structure in the nonmagnetic matrix. In fact, the $M_s^{7T}$ in a 3.2-nm-thick PdCoO$_2$ film was $3.4 \times 10^{-5}$ emu cm$^{-2}$, which corresponded to 2.6 $\mu_B$ per areal unit ($\sqrt{3}ab/2$, where $a$ and $b$ are lattice constants of PdCoO$_2$), and therefore much larger than the value deduced from the ARPES data (i.e. 0.59 $\mu_B$ per areal unit) [17].

The excess magnetic moment may have resulted from electron doping caused by band bending that extended over the insulating [CoO$_2$]$^-$ layer next to the top Pd surface, although the role of oxygen vacancies must also be considered. In contrast to the three-dimensional, antiferromagnetic PdCrO$_2$, the spins in the electron-doped CoO$_2$ layer are likely to fluctuate because of the low dimensionality and could thus be easily aligned by an external magnetic field. Given the parallel spin alignment at high field strength, a $CoO_2$ layer that is fully occupied with $Co^{2+}$ ions corresponds to 3 $\mu_B$ per areal unit for the high-spin state (Fig. 4(d), top) and 1 $\mu_B$ per areal unit for the low-spin state (Fig. 4(d), bottom). These values can account for the extra magnetization contribution of ~ 2 $\mu_B$ per areal unit at 7 T, which is added to the surface Pd contribution. The possible non-colinear magnetic moment in the electron-doped $CoO_2$ layer could couple with the conduction electrons in the adjacent Pd layer, giving rise to the $R_{UH}$ mapped to the temperature-$\mu_0 H$ plane that was shown in Fig. 4(c). In general, the dome-like structure of $R_{UH}$ in the temperature-$\mu_0 H$ plane was consistent with the behavior that has been observed in non-colinear spin systems [24-28]. The broken inversion symmetry at the surface might contribute to the generation of a non-trivial magnetic structure that can induce $R_{UH}$ via the Berry-phase mechanism [26,30]. Further studies examining these detailed magnetic structures are needed to understand the causes of $R_{UH}$ in $PdCoO_2$ thin films.

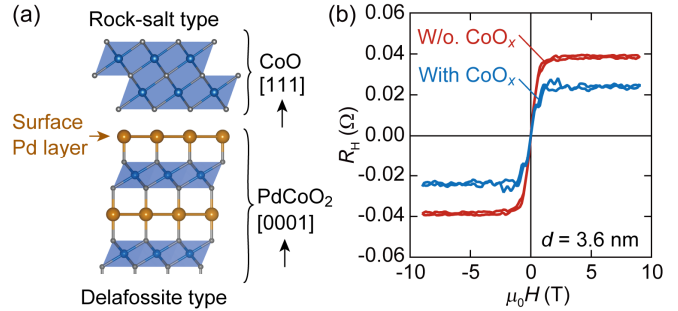

FIG. 3. Control of anomalous Hall effect by CoO$_x$ capping. (a) Schematic illustration of the rock-salt type CoO (111) heterostructure and the delafossite-type PdCoO$_2$ (0001). (b) $R_H$ versus $\mu_0 H$ for PdCoO$_2$ (3.6 nm) (red) and CoO$_x$ (5 nm) / PdCoO$_2$ (3.6 nm) (blue).

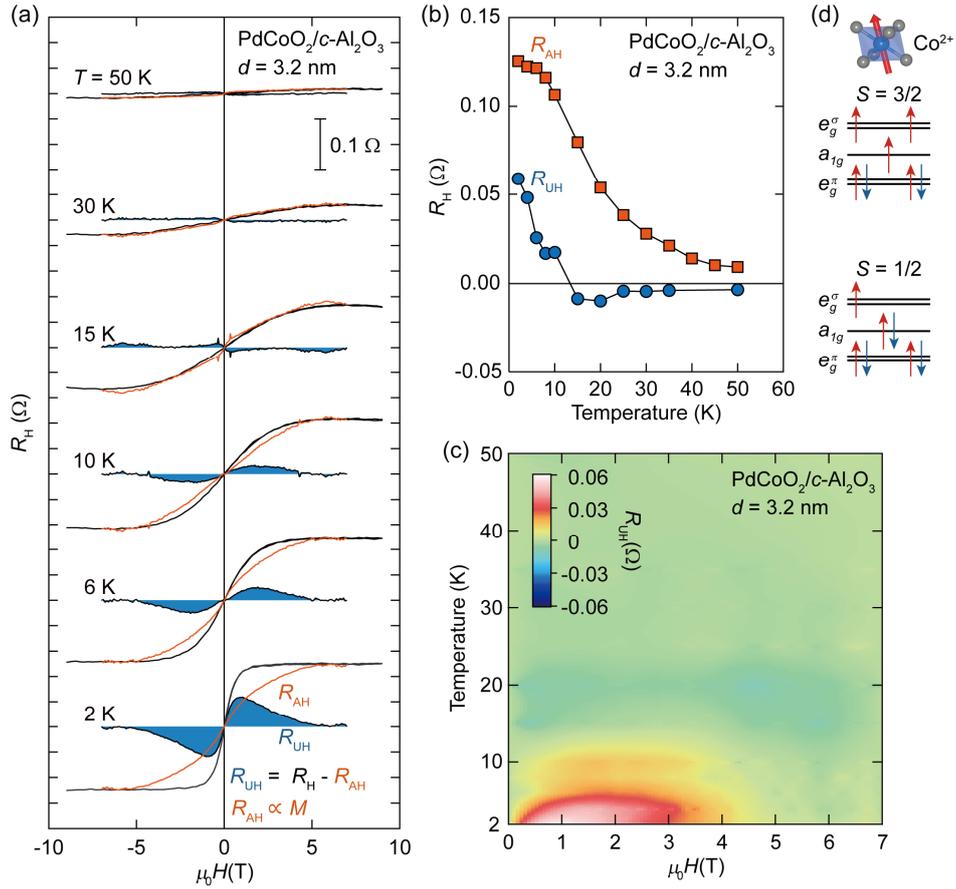

**FIG. 4.** Unconventional anomalous Hall effect of $PdCoO_2$ thin films. (a) Magnetic field ($\mu_0 H$) dependence of $R_H$, $R_{AH}$, and $R_{UH}$ for $d = 3.2$ nm at various temperatures. $R_{AH}$ was calculated using the magnetization curves. (b) Temperature dependence of $R_{AH}$ (red squares) and $R_{UH}$ (blue circles). (c) The $R_{UH}$ mapped as a function of temperature and magnetic field. (d) The electron configurations of $Co^{2+}$ in high-spin state (top) and in low-spin state (bottom).

In conclusion, we have studied the electronic properties at the polar surface of the layered, non-magnetic metal $PdCoO_2$ and have successfully detected surface ferromagnetism via the anomalous Hall effect using ultrathin $PdCoO_2$ films. We have confirmed that the origin of the ferromagnetism was from the surface of the films by systematically varying the thickness and termination layer of the films. ARPES measurements revealed Stoner-split surface Pd bands at the Pd-terminated surface. The unconventional, anomalous Hall resistance that was observed at low field strength was explained using the assumption that the $CoO_2$ layer was electron-doped, although direct observation of the Co valence state, e.g. using X-ray magnetic circular dichroism, is a future challenge. This work has demonstrated that using the thin-film approach [19,31] to study surface magnetism is highly effective because it drastically reduces bulk effects that can mask the surface properties. Revealing the $R_{UH}$ mechanism requires an understanding of the detailed nature of surface ferromagnetism, especially the spin interactions of doped $Co^{2+}$ ions and the role of broken inversion symmetry. The complex magnetic states at the surface, in combination with the hydrodynamic electron transport physics [13], make this layered oxide even more attractive for future research on spin-related transport phenomena.

We thank T. Nakamura and H. Oinuma for their assistance in the ARPES experiments. This work is a cooperative program (Proposal No. 18G0407) of the CRDAM-IMR, Tohoku University. This work is partly supported by a Grant-in-Aid for Specially Promoted Research (No. 25000003), a Grant-in-Aid for Scientific Research on Innovative Areas (No. 15H05853), a Grant-in-Aid for Scientific Research (A) (No. 15H02022), a Grant-in-Aid for Scientific Research (B) (No. 18H01821), a Grant-in-Aid for Early-Career Scientists (No. 18K14121) from the Japan Society for the Promotion of Science (JSPS), JST CREST (JPMJCR18T2, JPMJCR18T1), Mayekawa Houonkai Foundation, and Tanaka Foundation.